# *A COMPACT SYSTEM FOR REGISTERING AND PROJECTING STEREO VIEWS.*

José Joaquín Lunazzi, Samuel de Souza Moreira, Rafael Pedro da Silva, Guilherme Arruda Pedroso
Universidade Estadual de Campinas, Campinas-São Paulo-Brazil
lunazzi@ifi.unicamp.br

*keywords:* Stereoscopy, 3D imaging systems, 3D visualization, 3D adapters, Custom 3D systems, 3D projecting systems, 3D camera, Holographic screen, Gogless-less 3D viewing, Free viewing 3D.

Abstract

Mirror adapters can match a camera and a projector to the task of producing and visualizing a stereo pair. The imaging digital technology can benefit on the traditional optical technology of the last century to get more compact and affordable cameras and projectors. We describe a system which can be applied to different conventional projecting systems, as polarization an anaglyphic ones, useful at an amateur level but put in consideration for the industry. As an additional possibility, the same kind of projector can be adapted for goggles-less stereo viewing in a white light holographic screen whose functioning is clearly demonstrated.

## 1. Introduction

   In the XIX century photography was invented and soon later 3D stereo photography too: Charles Wheatstone in 1838 presented a stereoscope based in two mirrors(Simmons, 1996). Stereo imaging, the process of capturing two images of a scene from two close horizontally displaced views, is a process reaching two centuries of application but not in a continuously increasing pace. It had some moments of popularity and others in which remained within a few followers. To make a stereo pair it was necessary to use two cameras simultaneously or to displace one camera to take two viewpoints in sequence. In 1894 Theodore Brown patented a mirror system to make stereo pictures with a single camera based in two mirrors placed one aside the other, producing stereo photographic pairs which are known in history(Herbert, 1997). In the XIX century too some viewers were developed and improved, reaching a binocular lens system for individual use only. Although at the beginning of the XX century some adapters for single camera were developed (Kyle, 2004) the two-mirror stereo adapters were not popular and, by the middle of the XX century other systems were preferred: a four-mirror system equalizing the optical path was on the market, the synchronization of two cameras was possible with a double shutter cord, and double photographic cameras were made. Cinematography showed productions based on goggles from whom the most successful were based on the polarization principle. Many adapters were offered based in prisms acting like mirrors to obtain a stereo pair from a single camera. To the author's knowledge a similar system for projectors was not made, a double slide projector was not popular and it was usual to arrange two projectors aligned for imaging on the screen.  By the end of the century it use was very rare but the XXI century experiences a revival of the public interest due to technical improvements in the cinematography apparatus associated to digital techniques. More recently, an improved color filtering technique was developed by an interference light process that allows projecting on any screen without the need for metallizing its surface. Meanwhile, some techniques for eliminating the need of goggles are being researched, based on lenticular covers or parallax barrier techniques located over LCD displays. Domestic digital cameras for stereo registering are rare or very recent and the old adapters for single cameras were not revived. To the author's knowledge, a double projector for 3D digital photography or video was not made. Stereo is now-a-days more a technique for the people to receive, consume, than to produce. We present here a way to adapt digital cameras and projectors employing mirrors to achieve stereo photo and video capture and projection with a reduced budget. While the capture is a practical result, the projection system we made is only a primary prototype intended to divulge its possibilities.

**2. Capturing stereo images with a single camera with a two-mirror system**

Considering the objective of a camera to be approximated by a thin lens model, two viewpoints of an object can be registered resulting a division of the field of view, as in Fig. 1

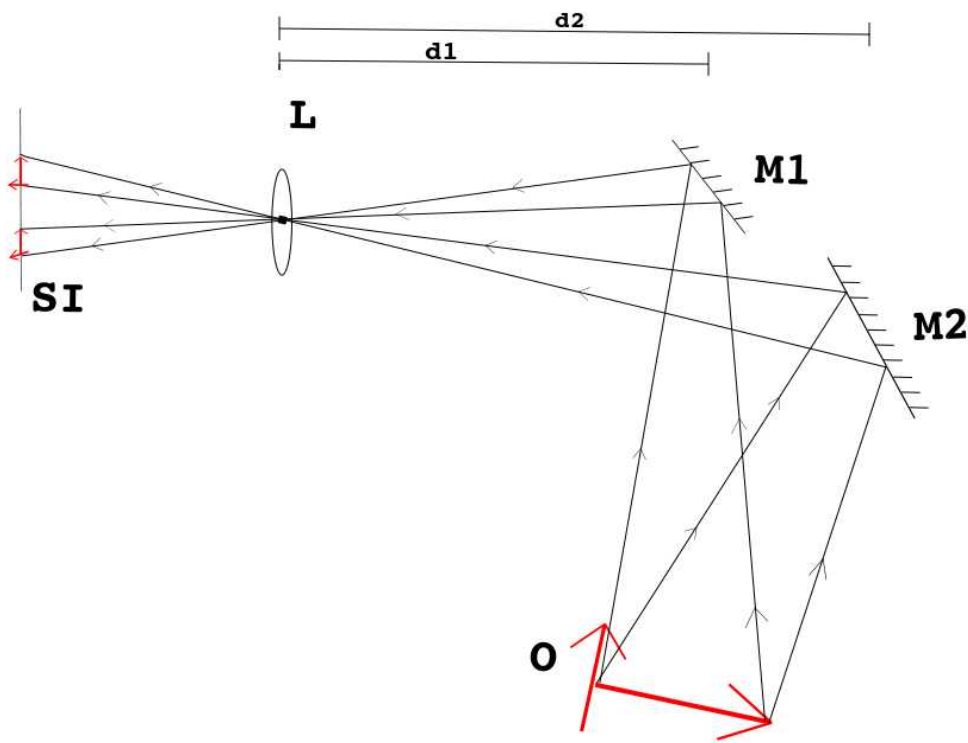

*Figure 1: Distribution of light in two viewpoints in a two-mirror system*

In one extreme of the object O, the point of an arrow, two rays impinging a different mirror each one (M1, M2) shows how two different views are projected on the stereo image pair SI . The extremes of the arrow were made to be coincident with the borders of the divided field of view. Both mirrors must reach the line corresponding to the optical axis of the lens L and have enough horizontal extension to correspond to the completion of the field of view. For a high resolution image the mirrors must be first surface coated to avoid astigmatism on the ray beams. The optical path is longer for one mirror's image and it has consequences on focusing matching and image size, so that the system is intended mainly for objects located at a minimal distance of 1 m under good illumination conditions to allow for reduced diaphragm aperture giving tolerance to focusing depth as corresponds to high f-numbers. The size difference can be as low as 1%, but it can also be corrected by digital means in a post-processing of the stereo pair. The division of the image sensor space changes the landscape format of an ordinary camera to portrait. This could be changed by a previous field compression if one cylindrical lens is located in the path of each stereo field after the light exits from the mirrors to the camera lens. First surface mirrors are the most appropiate because it avoids phantom images coming from reflection at the glass surface and astigmatism due to refraction on the glass thickness.

**3. Projecting stereo images with a single projector with a four-mirror system**

The field division can also be achieved with a projector. It seems to be a problem that could be solved as a reverse engineering one: just by reversing the sense of the light rays, the camera that registers two viewpoints of a scene in separate fields would project the two fields at the position of the scene, being received by a screen. But it is not so easy. Even for a camera, not always the design of its objective makes possible to have the mirrors close to its body. For some video projectors that we tried the situation appeared to be even worst, and we could only separate the fields at an intermediate distance between the projector and the screen. Even so, because we know that the situation could be solved with proper optical design on the curvatures and refractive indexes of the objective's lenses, we developed the system as a demonstrative proposal to consider making a compact system. Making a two-mirror system becomes inadequate because of the limited tolerance to focusing when two distances are involved. One arm makes a different path than the other and, due to the need of high luminousity on the objective, large apertures are necessary and then focusing becomes strict. So that a four mirror system must be employed, in a symmetrical path situation shown in Fig. 3. In that ray tracing scheme letters **d** indicates distances involved, **P** indicates the projector, M indicates mirrors and **S** the screen.

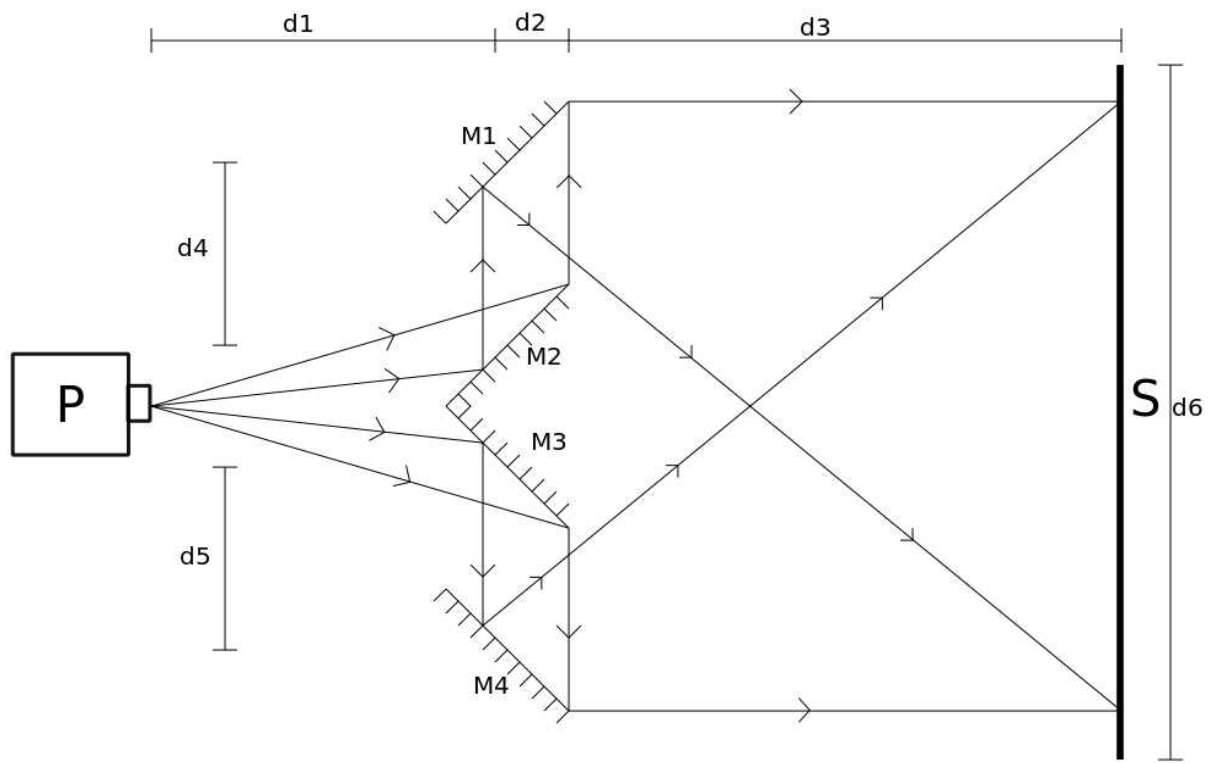

*Figure 3: Ray tracing scheme for a projector overlapping the two half of its field on a screen*

**4. Experimental details for the camera**

To capture a scene a Sony Handycam HDR-CX100 AVCHD HD Camcorder was employed and two ordinary plane mirrors fixed on an aluminum support plate. The one closer to the camera was 3 cm large and 9 cm high, located at 5.5 cm distance $d_1$ from the camera, while the second one was 7 cm large and 10.3 cm high and at 8.5 cm distance from the camera (Fig. 1). Calibration of the angular position for both was made by fixing a reference small object at the working distance and then making its image to appear centered on both semi-fields of the viewer camera screen. The lateral distance of the mirrors must correspond to the desired degree of horizontal parallax. It is not difficult to work around a value of 6 cm corresponding to binocular parallax or to design a mount were it can be adapted to different situations according to the scene and object distance. The resulting image is shown in Fig.2, already digitally reversed from left to right to make it more natural.

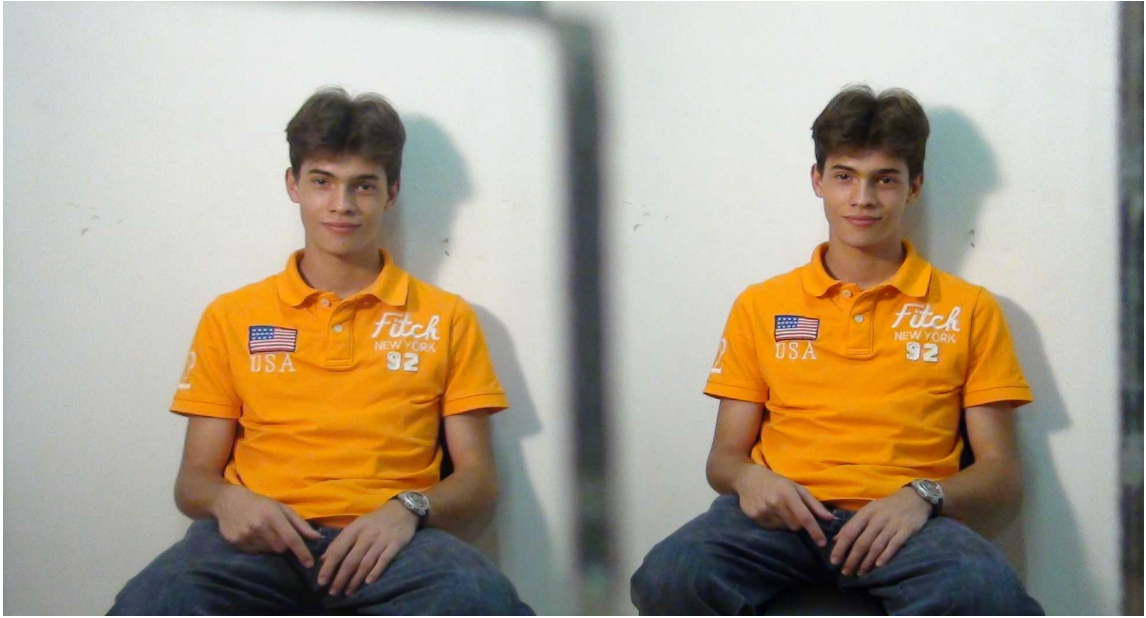

*Figure 2: An stereo pair obtained from a single camera with two mirror field division*

From the side-to-side stereo pair digital processing can render the 3D image to the desired 3D format. Besides the professional standards of common use, the domestic manipulation can be made to obtain an anaglyph, as seen in Fig. 3. Open source software such as (GIMP, 2012) can be employed which is available for many operating systems.

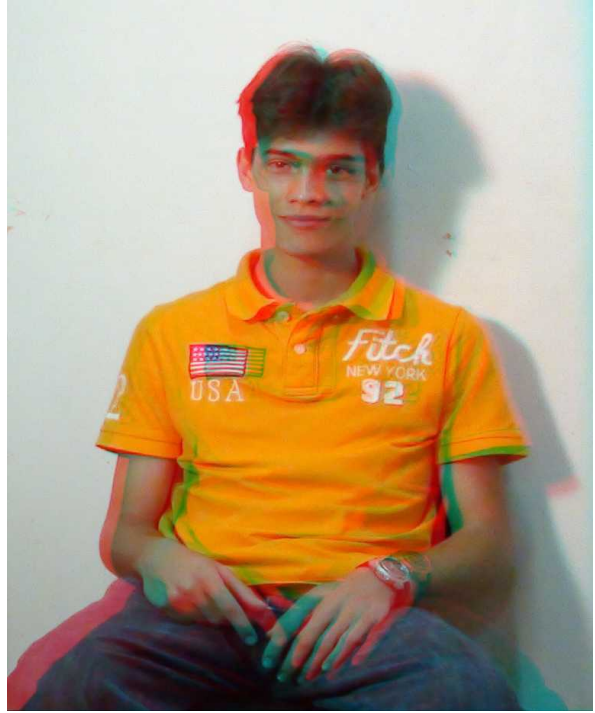

*Figure 3: an anaglyph made from a single camera stereo pair*

The only need was to eliminate the blue and green color channels for the left-eye view and the red color channel for the right-eye view.

## 5. Experimental details for the projector

When making the projector system a ViewSonic PJ503D 1,500 lumens projector was adapted by replacing its short focal lens by a 1:2.8 85 mm IEC slide projector objective in order to find a place where to intercept the beam with reduced size mirrors. The beam splitter was then made by using common mirrors: four elements 3 mm thick and 18 cm x 18 cm large were mounted in a 72 cm large mount at 45 degrees with the projector beam, located at 1.70 cm from the projector (Fig.4). At a shorter distance the separation of the left-hand side image and the right-hand side one was not well defined enough. The angle adjustment for the mirrors was made with ease by deformation of Blue-Tack plastic material.

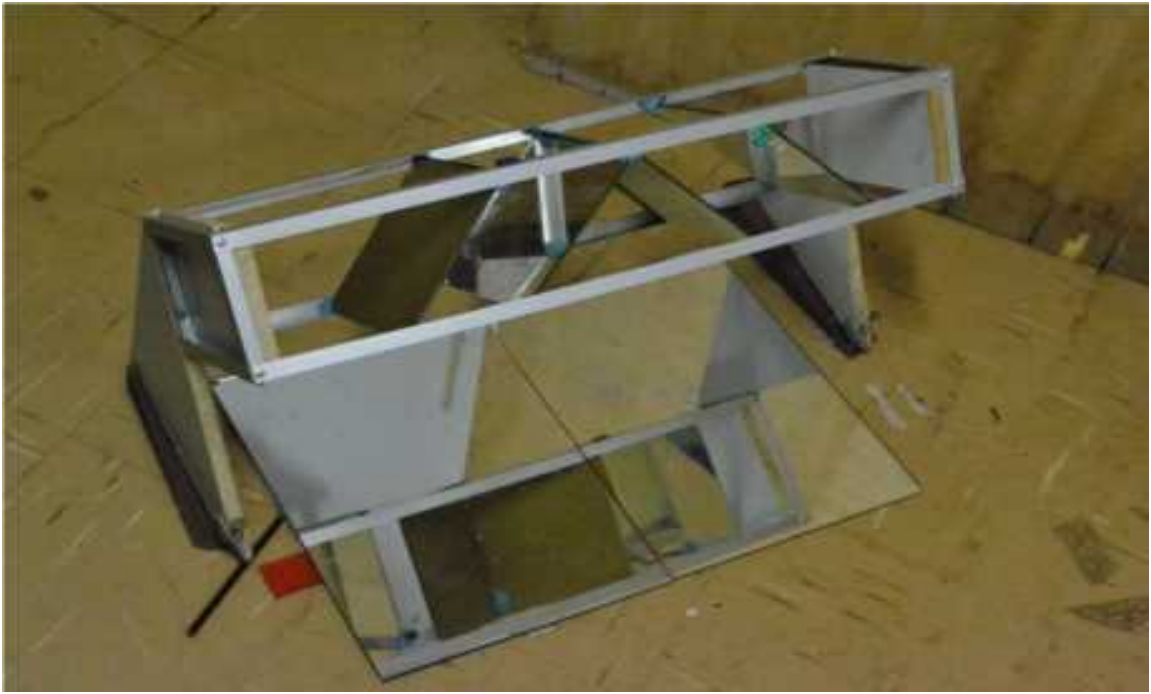

*Figure 4: A four mirror beam splitter system. The two mirrors on the ground are not part of it.*

The distance between the projector and the screen was 5.7 m and the size of the projection 53 cm wide and 70 cm high. 18 cm x 13 cm filters were located at the beam spliltter, in two-color acetate for anaglyphic projection in an ordinary screen or a brighter image when filtering with linear polarizing material to project on a metallized screen. Só that the side-by-side photo or video image obtained with the two-mirror single stereo camera can be directly projected without the need of any digital processing to adapt its format. Although with reduced quality due to non-properly designed optics and not first surface mirrors, the result was good enough to give 3D realm (Fig.5).

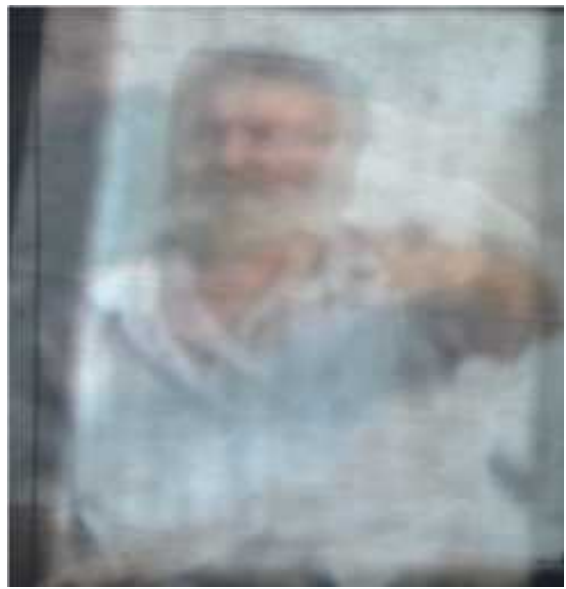

**Figure 5: Superposed stereo pair polarized image obtained at the screen.**

## 6. A projector system for a holographic screen

Another application can be made with the four-mirror system stereo projector if the angle of incidence at the screen matches the angle for binocular viewing of the observer. This was our case and a holographic screen 50 cm large and 61 cm high(Lunazzi, 2009, 2008) received the beams impinging at 45 degrees downward to be concentrated at 1.1 m after the screen. Figure 6 shows how each eye of the observer sees each separate view of the stereo pair because the light from the image illuminates his face at the eyes.

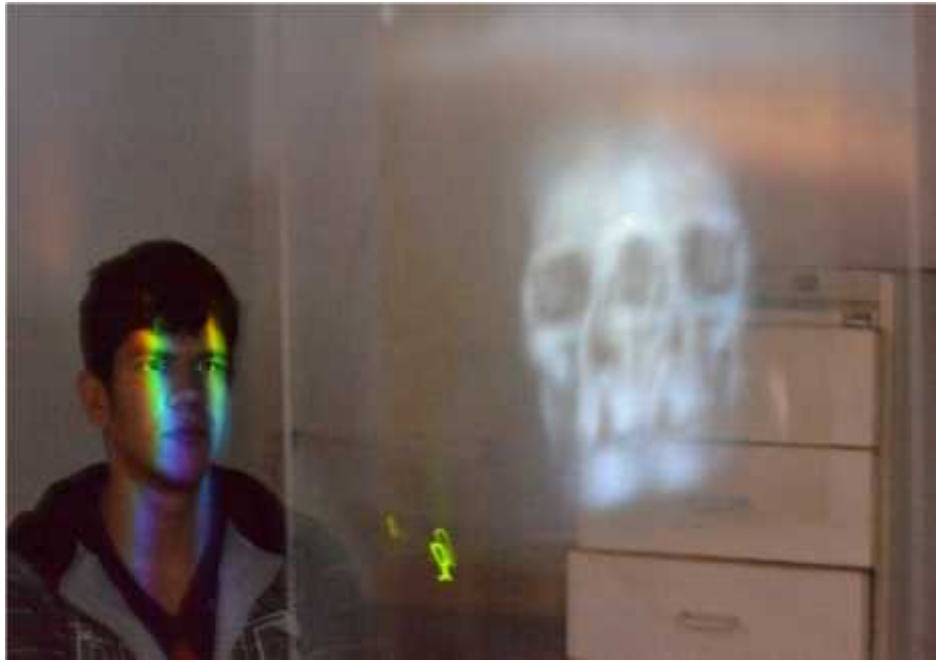

**Figure 6: Stereo pair of a skull being superposed at the screen and separated at the observer's eyes.**

The screen case shows a critical situation with reduced lateral field for the viewer, but this is just for demonstrating the principle. A screen can be made giving a lateral dispersion wide enough as to give more than 6 cm of lateral field(Magalhães, 2012).

## 7. Conclusions

An amateur low-cost equipment for making projection stereo photography was demonstrated that could be profitable at industrial level if properly designed. The registering part of it can be made to have 3D video and photography on the computer or on the projector system. The projector system can also be useful for gogless-less stereo viewing by means of holographic screens.

## 8. Acknowledgements

The Pro-Reitoria de Graduação-PRP of Campinas State University-UNICAMP, is acknowledged for the PROFIS fellowship given to students G.A. Pedroso, the Pro-Reitoria de Pesquisa, PRP-UNICAMP, is acknowledged for the PIC Jr fellowship given to S.S. Moreira, R.P. da Silva, C.L.D. Bargas, for the SAE fellowship given to students A.R. Da Costa, V.E. Ares, C.A.I. Canhassi, C.S. Rennó, all of them collaborators in the project. And for resources coming from the project "Ciência e Arte nas Férias". The Pro-Reitoria de Extensão, PREAC-UNICAMP, is acknowledged for sustained support in projects that benefit directly to the public and the Coordenadoria de Aperfeiçoamento do Ensino Superior-CAPES of the Ministry of Education is acknowledged for resources helping to acquire a multimidia projector and holographic material. The ideia and project are from J.J. Lunazzi while S.S. Moreira, R.P. Da Silva and G.A. Pedroso collaborated in making the image capture and projection systems.

## 10. Later comments on march 2013
The author knew that the recent high-quality projector 4K from Sony does employs e field beam splitter that could be considered as the one proposed. So that the idea of this article seems to be useful to divulgate.